\documentclass[10pt]{iopart}

\usepackage{graphicx}
\usepackage{bm}
\usepackage{braket}
\usepackage{multirow}
\usepackage{rotating}
\usepackage{todonotes}
\usepackage{array}
\usepackage{color}
\usepackage{feynmf}
\usepackage{hyperref}
\usepackage{enumerate}
\begin{document}

\title[Computational design of metal-supported molecular switches]{Computational design of metal-supported molecular switches: Transient ion formation during light- and electron-induced isomerisation of azobenzene}
\author{Reinhard J. Maurer}
\address{Department of Chemistry, University of Warwick, Gibbet Hill Road, CV4 7AL, Coventry, United Kingdom}
\address{Department Chemie, Technische Universit{\"a}t M{\"u}nchen, Lichtenbergstr. 4, D-85748 Garching, Germany}
\ead{r.maurer@warwick.ac.uk}
\author{Karsten Reuter}
\address{Department Chemie, Technische Universit{\"a}t M{\"u}nchen, Lichtenbergstr. 4, D-85748 Garching, Germany}

\vspace{10pt}
\begin{indented}
\item[]August 2018
\end{indented}

\begin{abstract}

In molecular nanotechnology, a single molecule is envisioned to act as the basic building block of electronic devices. Such devices may be of special interest for organic photovoltaics, data storage, and smart materials. However, more often than not the molecular function is quenched upon contact with a conducting support. Trial-and-error-based decoupling strategies via molecular functionalisation and change of substrate have in many instances proven to yield unpredictable results. The adsorbate-substrate interactions that govern the function can be understood with the help of first-principles simulation. Employing dispersion-corrected Density-Functional Theory (DFT) and linear expansion Delta-Self-Consistent-Field DFT, the electronic structure of a prototypical surface-adsorbed functional molecule, namely azobenzene adsorbed to (111) single crystal facets of copper, silver and gold, is investigated and the main reasons for the loss or survival of the switching function upon adsorption are identified.  The light-induced switching ability of a functionalised derivative of azobenzene on Au(111) and azobenzene on Ag(111) and Au(111) is assessed based on the excited-state potential energy landscapes of their transient molecular ions, which are believed to be the main intermediates of the experimentally observed isomerisation reaction. We provide a rationalisation of the experimentally observed function or lack thereof that connects to the underlying chemistry of the metal-surface interaction and provides insights into general design strategies for complex light-driven reactions at metal surfaces.


\end{abstract}

%

\submitto{\JPCM}
%
%
\ioptwocol

\section{Introduction}
\label{intro}

A key prerequisite to develop applications of molecular nanotechnology is the ability to selectively and reversibly change measurable properties of single molecules. Materials and techniques which enable this are important for molecular-level optoelectronic device integration,\cite{Cuevas2010} molecular circuits,~\cite{Mativetsky2008} ultrasensitive molecular sensors,~\cite{Joshi2014} and plasmonic reaction control at nano-structured interfaces.~\cite{Xie2018}

Single molecules that can be reversibly switched between two or more distinguishable states, so-called molecular switches, provide an interesting playground for molecular device design.~\cite{Feringa} The changes between these states can be in the form of discrete modifications to the molecular conformation, the chemical state, the magnetic state, or other measurable molecular properties. Embedded in a matrix~\cite{Ikeda1995}, in porous metal-organic frameworks~\cite{Wang2015} or adsorbed at well-defined surfaces \cite{Katsonis2006}, such molecules are envisioned to be switchable individually or in domains, and the corresponding changes have to be detectable in terms of simple electric or spectroscopic signatures.  In practice, this is an intricate task, because the function of molecular switches depends on a delicate balance of different intra- and intermolecular interactions, but also on the interactions with the environment. Conducting electrodes, such as metal substrates are of special interest as support due to the direct contact of the molecule to a conductor and the prospects of controlling function via an applied voltage or via introducing novel functionality due to electronic coupling. However, the strong molecule-metal coupling can readily lead to a loss of function -- if the intermediate electronic states are quenched, the reaction pathway is hindered, or more fundamentally, the basic stability of the equilibrium geometries changes.~\cite{Maurer2012,Titov2015}

Although clear-cut design strategies are still lacking, a large variety of molecular switches is currently investigated for their functionality. Explorative substrate-variation and molecule functionalisation have led to many switchable metal-organic interfaces  including molecules that undergo changes due to chemical reactions~\cite{Schulze2012}, conformational changes~\cite{Comstock2005}, switching between chemisorbed and physisorbed states~\cite{Liu2013}, or magnetic spin transitions \cite{Venkataramani2011, Wackerlin2010}. The means to induce such switching range from light irradiation to thermal activation~\cite{Hagen2007} and inelastic electron tunnelling~\cite{Choi2006}. These fundamental studies have led to fascinating prototype applications such as molecular motors or machines~\cite{Browne2006,Chiang2012, Coskun2012} and optically contractable polymers~\cite{Hugel2002a}. 

\begin{figure}
	\centering\includegraphics[width=3.3in]{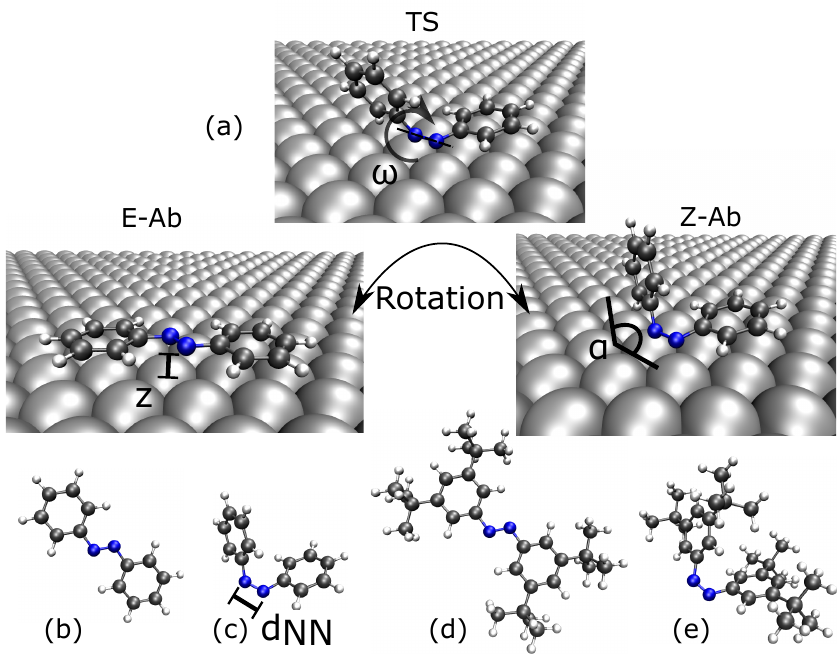}
	\caption{\label{fig-params} (a) Schematic view of the relevant degrees of freedom for the on-surface isomerisation between E and Z-Ab on a Ag(111) surface via a transition state (TS) along the dihedral rotation of the central di-nitrogen bond. N atoms are shown in blue, C atoms in grey, and H atoms in white. The Ag atoms are shown as van-der-Waals spheres. Relevant geometry parameters are defined: $z$ refers to the vertical adsorption height of the N atoms, $\omega$ refers to the dihedral angle around the di-nitrogen bond (CNNC), $\alpha$ refers to the CNN bending angles. The bottom panels show equilibrium gas-phase structures of isolated (b) E-Azobenzene (E-Ab), (c) Z-Ab, (d) E-3,3'5,5'-tetra-\emph{tert}-butyl-Azobenzene (E-TBA), and (e) Z-TBA.}
\end{figure}

Azobenzene (Ab) and its derivatives represent one of the best studied classes of molecular switches. Its ability to change between an E and a Z conformation around the central N-N double bond (cf. Fig. \ref{fig-params}) upon light irradiation depends not only on an efficient excited-state reaction channel, but more fundamentally on the sheer fact that two meta-stable states coexist in the ground state and are fairly stable at the experimental conditions. Owing to time-resolved photoexcitation experiments and quantum dynamics simulations, the electronic and atomistic details of the gas phase photoswitching mechanism are, by now, quite well understood~\cite{Pederzoli2011,Gamez2012, weingart2011}. For a review of the literature see Ref. \cite{Bandara2012} or \cite{Maurer_Thesis2014} page 37.

The persistence or loss of switching function of Ab and its derivatives directly adsorbed to noble metal surfaces, such as close-packed facets of coinage metals, has been intensively studied  \cite{Comstock2005,Kirakosian2005,Henzl2006,alemani2008, Wolf2009, Henzl2012,Morgenstern2011}; no light-induced switching of pure Ab on Ag(111) or Au(111) has been observed to date.~\cite{comstock2007} The lack of photo-induced function has been attributed to three main factors: overly strong coupling to the surface, quenching of the excited states involved in the isomerisation, and steric hindrance due to strong dispersion interactions between the phenyl rings of the molecule and the underlying surface.~\cite{mcnellis2010a} However, successful switching of Ab on Au(111) via resonant electron tunnelling through a Scanning Tunnelling Microscopy (STM) tip~\cite{Choi2006}, suggests that there are additional factors at play when comparing different substrates. In all experimental studies, successful switching has been verified either by significant  changes to spectral features, such as Two-Photon-Photoemission (2PPE) resonances, or by visual inspection of STM topograph changes. 

Molecular functionalisation strategies to decouple the photochromic azo-moiety from the surface follow the basic ideas of either geometrically decoupling the central di-nitrogen bridge from the surface with spacer groups~\cite{Alemani2006,comstock2007,Hagen2007},  introducing a modified switching process via strongly chemically active functional groups~\cite{Henzl2006, Henzl2012,Morgenstern2011}, or via combination of Ab units into sterically constrained tripods.~\cite{Scheil2016}  Comstock \emph{et al.}  tested the former by analysing the switching behaviour of Ab and Ab substituted with \emph{tert}.-butyl groups at the two phenyl \emph{para} positions and the four \emph{meta} positions (DBA and TBA, respectively) when adsorbed to Au(111).~\cite{comstock2007} The authors were able to detect a switched state of TBA after 1 hour of irradiation with UV light, whereas no isomerisation was observed for Ab and DBA. The reason for a successful photo-induced isomerisation of TBA in contrast to DBA and Ab has been related to the electronic decoupling of the central nitrogen atoms,  therewith reduced substrate effects on the excited-state lifetime, and the optical absorption properties. We note that all three molecules represent functioning molecular switches in gas phase or solvent and could be switched by an STM tip when adsorbed at a Au(111) surface~\cite{Choi2006,Alemani2006,comstock2007}.

The isomerisation mechanism of metal-mounted TBA is believed to be strongly modified compared to the gas phase. The change in electronic structure of the adsorbate can be verified by STM topographs~\cite{comstock2008}, 2PPE signatures of the molecular resonances~\cite{Hagen2008}, and changes in the work function.~\cite{Hagen2007} The photo-switched state can be  identified as the molecular Z-TBA state by structural predictions from Angular Resolved Near-Edge X-Ray Absorption Fine-Structure (NEXAFS) experiments~\cite{Schmidt2010}. In contrast to the gas phase or solvent situation, a photo-induced back-reaction from this Z-TBA state to E-TBA is much less efficient and does not happen in the visible regime, but rather with photon energies in the same energy regime as the E-to-Z transition.~\cite{comstock2008} A corresponding drop in switching efficiency can thus already be understood in terms of simultaneously induced switching in both directions, as is supported by the reduced E/Z ratio at the photostationary state. Wolf and Tegeder have proposed a mechanism of cationic resonance formation in E-to-Z photo-isomerisation of TBA that is in agreement with experimental observations.~\cite{Wolf2009} Light irradiation generates electron-hole pair excitations in the metal. The generated holes rapidly diffuse to the upper edge of the metal d-bands. Energetic overlap between the d-bands and occupied molecular frontier orbitals, such as the highest occupied molecular orbital (HOMO) will enable a transient cation formation in the molecule, which subsequently drives the isomerisation.~\cite{Hagen2008, Tegeder2007} The prevalent assumption for the ensuing molecular motion is that, although the coupling to the metallic degrees of freedom will limit the lifetime of this state to  tens to hundreds of femtoseconds,~\cite{campillo2000} the kinetic energy gained by the motion on the excited-state PES might suffice to enable the transition to the Z-Ab state. This is the so-called Menzel-Gomer-Redhead model~\cite{Menzel1964,Redhead1964}, which has been proposed in the context of photo-induced desorption processes~\cite{Zhou1991, Hertel1995, Guo1999}. Not much is known about the atomic degrees of freedom involved in surface-mounted TBA switching though. On the basis of STM topographs and chiral mapping, it has been argued that the isomerisation must follow a predominantly planar inversion of the central CNN bending angle $\alpha$ (\emph{cf.} definition in Fig.~\ref{fig-params}) rather than the rotational mechanism that is believed to be dominant in gas phase.~\cite{Comstock2010}

First-principles modelling of such light-induced molecular adsorbate reactions and involved excited states can in principle shed more light into the dynamics and the mechanism. Presently, this is extremely challenging. This relates to the large system size, the concomitant high computational expense, and the difficulty of describing electronic charge-transfer excitations at the molecule-metal interface. An important prerequisite for an \emph{ab initio} simulation of the excited-state dynamics involved in photo-isomerisation is a detailed understanding of the adsorption geometries of the ground-state equilibrium structures and the ability to access electronic excited states of molecular adsorbates at metal surfaces. To this end, the recent emergence of predictive-quality dispersion-corrected Density Functional approximations for metal-organic interfaces~\cite{Ruiz2012, Maurer2016Review} and efficient, approximate methods to calculate electronic excited states, such as linear expansion Delta-Self-Consistent-Field DFT ($\Delta$SCF-DFT),~\cite{Maurer2013,gavnholt2008} is providing the necessary tools to study the fundamental structural and electronic parameters that control light- or electron-driven Ab isomerisation at metal surfaces in more detail.

In this work, we use these tools to study the structural and electronic properties at the metal-organic interface that  govern the switching ability of metal-adsorbed functional molecules. We do this for the case of coinage-metal adsorbed Ab, for which we previously were able to rationalise the loss of function on Ag(111) as a loss of ground state bi-stability~\cite{Maurer2012}. By analysing the effects of substrate variation and molecule functionalisation for coinage-metal-adsorbed Ab and TBA on the ground- and excited-state potential energy landscapes of the adsorbed species, we connect how chemical and electrostatic molecule-surface interactions shape these energy landscapes and provide the basis for molecular switching or the lack thereof.

\section{Computational Methods}
\label{methods}

\subsection{Density Functional Theory calculations}

All DFT calculations have been performed with the pseudopotential plane wave code CASTEP version 6.0.1~\cite{Clark2005,Payne1992} using the legacy MaterialsStudio ultrasoft pseudopotentials (USPPs)~\cite{Vanderbilt1990}. The exchange-correlation functional due to Perdew, Burke, and Ernzerhof (PBE)~\cite{Perdew1996} has been used throughout. We account for dispersion interactions using the screened Tkatchenko-Scheffler pairwise dispersion correction method (vdW$^{\mathrm{surf}}$).~\cite{Ruiz2012} This method describes the screening of van-der-Waals interactions in the metal substrate by renormalisation of the pairwise C$^6$ coefficients of the metal. It has, in the past, been shown to provide accurate adsorption geometries at a moderate overestimation of the adsorption energies.~\cite{Mercurio2013, Maurer2016, Maurer2016Review} Higher-order dispersion correction techniques, such as the Many Body Dispersion (MBD) method exist to correct for this overbinding,~\cite{Tkatchenko2012,Maurer2015}, but have not been available for geometry optimisations at the time of this study. For all structures, the dispersion interactions between periodic images of the adsorbed molecules have been switched off to simulate a low-coverage situation. 

The employed computational set-up was previously described with details on careful convergence tests~\cite{Maurer2012,McNellis2009}. In short, all calculations were performed on (111)-oriented bulk-truncated 4-layer surface slabs of Cu, Ag, and Au with 350~eV or 450~eV plane wave cutoff for Azobenzene and 3,3',5,5'-tetra-\emph{tert.}-butyl-Azobenzene (TBA), respectively. The top-most two layers of the substrate slabs were allowed to relax.  The  PBE optimised lattice constants for Cu, Ag, and Au, are 3.62~\AA{}, 4.14~\AA{}, and 4.19~\AA{}, respectively. The vacuum region above the surface slab was chosen to exceed 20~\AA{} in all cases. We obtained optimized geometries using the delocalized internal coordinate optimization algorithm in CASTEP~\cite{Andzelm2001} with a maximum atomic force threshold of 25~meV/\AA{}. Minimum energy paths were calculated by constrained optimization where the value of the central dihedral angle $\omega$ was fixed while all other degrees of freedom of the molecule were allowed to relax. The transition states reported in Table 1 have been calculated with the quadratic synchronous transit method.~\cite{Govind2003}

The results in Fig.~\ref{fig-results-ab-rotation} have been obtained for (6$\times$4), (6$\times$3), (6$\times$3) surface unit cells containing a single Ab molecule on Cu, Ag, and Au, respectively. 
TBA molecules have been studied in (6$\times$4) and (6$\times$5) surface unit cells on Ag and Au, respectively. Calculations of the gas phase molecules have been performed in 40$\times$40$\times$40~\AA{} and 50$\times$50$\times$50~\AA{} rectangular supercells for Ab and TBA, respectively, with the electronic structure calculated only at the $\Gamma$-point of the first Brillouin zone.

\subsection{Modelling molecule-surface charge-transfer excitations with linear expansion $\Delta$-Self-Consistent-Field-DFT (le$\Delta$SCF-DFT) }

The excited state results described in section \ref{results-excited-state} have been obtained with the linear expansion $\Delta$-Self-Consistent-Field DFT method (le$\Delta$SCF-DFT), which we have implemented in CASTEP 6.0.1 and which is available in the latest CASTEP 18.1 release.~\cite{Maurer2013} Conventional $\Delta$SCF-DFT, where the occupation of a single Kohn-Sham (KS) state is constrained to model an excited state, would not be able to correctly capture the charge-transfer excitation between an adsorbed molecule and the metal surface in the presence of electronic hybridisation between the two. The le$\Delta$SCF-DFT method addresses this issue by construction of a resonance orbital. This is achieved via projection of the wave function of a molecular orbital of the free molecule $\ket{\phi_c}$ onto the KS states of the surface slab containing the molecule: 
\begin{center}\begin{eqnarray}
\ket{\tilde{\psi_c^{\mathbf{k}}}} = \sum_i^{\mathrm{states}} \ket{\psi_i^{\mathbf{k}}} \braket{\psi_i^{\mathbf{k}}|\phi_c}
\end{eqnarray} \end{center}

Upon construction of this resonance state $\ket{\tilde{\psi_c^{\mathbf{k}}}}$, all remaining KS states are orthonormalised with respect to it. The orbital occupations are then distributed and constrained to model the envisioned excitation. This procedure is repeated during each consecutive Self-Consistent-Field (SCF) step until convergence is reached.  

In Ref.~\cite{Maurer2013}, we have shown that this method can be used to model intramolecular excitations of electrons from occupied to unoccupied adsorbate states, such as the Ab $\pi\rightarrow\pi^*$ (HOMO$\rightarrow$LUMO or S2) and the n$\rightarrow\pi^*$ (HOMO-1$\rightarrow$LUMO or S2) excitations. However, also charge-transfer excitations between the molecule and the surface can be modelled. In the case of a hole transfer from the substrate to the molecule, a so-called cationic resonance (HOMO$\rightarrow\epsilon_F$ or $h^{+}_{\mathrm{HOMO}}$), we constrain the occupation of the HOMO to be one electron less than in the ground state. The missing electron is balanced by adjusting the Fermi energy accordingly, which has a minimal effect on the electronic structure in the case of a large metal surface slab. In the case of an electron transfer from the substrate to the molecule, a so-called anionic resonance ($\epsilon_F\rightarrow$LUMO or $e^{-}_{\mathrm{LUMO}}$), one electron is taken from the Fermi level and added to the LUMO of the molecule.

\begin{figure}
	\centering\includegraphics[width=3.3in]{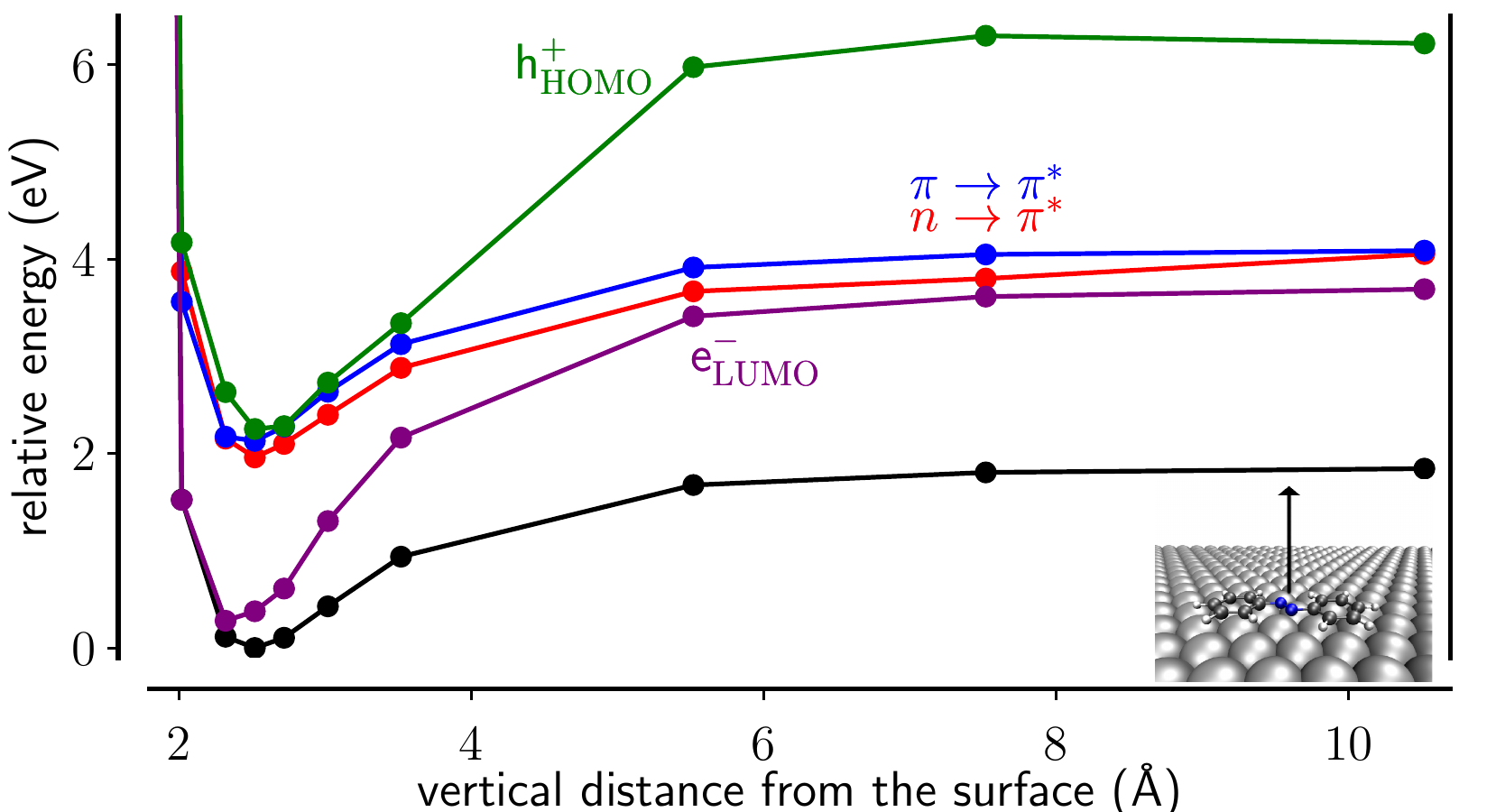}
	\caption[Binding energy curve and excited state energies of Azobenzene on Ag(111)]{\label{fig-ledscf-bindingcurve} Ab adsorbed to Ag(111): Relative energy with respect to the ground-state equilibrium adsorption geometry as a function of the vertical adsorption height. Points at 10.5~\AA{} distance represent a molecular free-standing overlayer. Shown are the ground state energy (black curve), the first (n$\rightarrow\pi^*$, red curve), and second ($\pi\rightarrow\pi^*$, blue curve) intramolecular excited states as well as charge-transfer excitations, where an electron is added to the LUMO of the molecule (e$^-_{\mathrm{LUMO}}$, purple curve) or removed from the HOMO of the molecule (h$^+_{\mathrm{HOMO}}$, green curve). The small model in the inset on the bottom right depicts the dissociation direction orthogonal to the surface.}
\end{figure}

Figure~\ref{fig-ledscf-bindingcurve} presents the ground- and excited-state energies of E-Ab on Ag(111) as a function of the vertical distance from the surface. The ground-state energetics and geometry are calculated with DFT+vdw$^{\mathrm{surf}}$. At every geometry four different le$\Delta$SCF-DFT calculations have been calculated to yield the pictured excited-state energies; two of them are charge-neutral intramolecular excitations, the other two are molecule-surface charge-transfer excitations. Far away from the surface, at distances above 10~\AA{}, the intramolecular excitations approach the limit of the isolated molecule (gas phase S1(n$\rightarrow\pi^*$): 2.27~eV , S2($\pi\rightarrow\pi^*$): 2.75~eV ).~\cite{Maurer2011} More precisely, the values approach the limit of an infinite free-standing molecular layer in the periodic arrangement of the employed supercell geometry (S1: 2.21~eV, S2: 2.24~eV).  The corresponding substrate-mediated excitations, h$^+_{\mathrm{HOMO}}$ and e$^-_{\mathrm{LUMO}}$, should, in the limit of infinite molecule-surface separation, approach the corresponding Ionization Potential (IP) and Electron Affinity (EA) of the molecule as predicted by the PBE functional (IP:~7.82~eV, EA:~1.06~eV \cite{McNellis2009a}). The corresponding value for the IP above 6~\AA\ distance from the surface is the difference between the  h$^+_{\mathrm{HOMO}}$  state  and the ground state (4.37~eV). The value of the EA can be calculated as the difference of the work function (including the potential drop due to the free standing overlayer this amounts to 4.21~eV \cite{McNellis2009a}) and the e$^-_{\mathrm{LUMO}}$ state and thereby amounts to 2.36~eV. The substrate-mediated excitation energies are very sensitive to the distance from the substrate and to the interactions with the neighbouring Ab images, as well as the periodic image of the substrate above~\cite{comment1}. This sensitivity stems from the excess charge on the molecule interacting via long-range Coulomb forces with the electron density of the substrate and the periodic images of the molecule. 

When approaching the surface, the intramolecular excitation energies do not change dramatically, whereas e$^-_{\mathrm{LUMO}}$ and h$^+_{\mathrm{HOMO}}$ strongly shift. Close to the equilibrium geometry the anionic resonance state is only 0.38~eV above the ground state and the cationic resonance state is at 2.52~eV above the ground state. This arises from the Coulomb interaction between the localized charge on the molecule and the substrate, which steadily builds up an image charge as the molecule approaches. This difference in the renormalization behaviour of intramolecular excitations and charge-transfer excitations is well known. It can be described at the level of many-body perturbation theory~\cite{Neaton2006, Thygesen2009, Garcia-Lastra2011}, whereas linear response time-dependent DFT based on semi-local functionals fails to capture this effect. It is remarkable that by employing such an approximate scheme as le$\Delta$SCF, the main physical effects of substrate polarization and excited-state renormalization can be qualitatively captured. We therefore feel encouraged to apply this method to study the excited state landscape of charge-transfer states during the Ab and TBA on-surface isomerisation in section~\ref{results-excited-state}.

\section{Results}
\label{results}

\subsection{The role of substrate electronegativity}
\label{results-azobenzene}


The most basic prerequisite to retain the isomerisation ability of Ab-based molecular switches upon metal-surface-adsorption is that the interaction with the substrate does not destroy the existence of the two meta-stable conformers in the ground state, namely E and Z-Ab (see Fig.~\ref{fig-params}). Strong reduction or removal of the ground-state barrier or changes in the relative stability of these two states would affect the thermal stability of a Z-Ab state. As such, any viable metal substrate to support functioning metal-mounted molecular switches  needs to have a weak interaction with the molecule that, ideally, affects all molecular conformations similarly as not to modify the energetic landscape along the reaction coordinate of isomerisation. Unreactive noble metals such as Ag(111) or Au(111) therefore come to mind as natural substrate choices.

\begin{figure}[bh]
	\centering\includegraphics[width=3.3in]{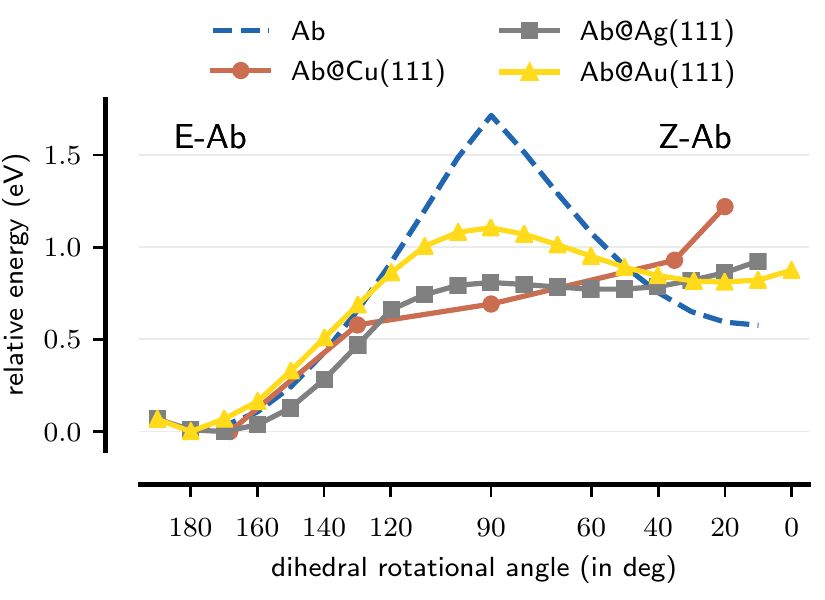}
	\caption{\label{fig-results-ab-rotation} Ground state minimum energy path from E-Ab to Z-Ab along the central dihedral rotation angle $\omega$ for Ab in gasphase (blue line, dashed) and adsorbed at Cu(111) (brown line, circles), Ag(111) (grey line, squares), and Au(111) (yellow line, triangles) surfaces. }
\end{figure}

In previous work, we have shown that Ab adsorption on these two surfaces does, in fact, still modify the energetic landscape along one of the three traditionally discussed pathways for the isomerisation reaction.~\cite{Maurer2012} These relevant reaction coordinates include the rotation around the central dihedral angle $\omega$, an inversion around one of the two central bending angles $\alpha$, and a symmetric inversion around the two central bending angles leading to a linearisation of the CNNC group. Various different pieces of evidence for the prevalence of one of these mechanisms over the others have been given for different enviroments.~\cite{weingart2011, Pederzoli2011, Titov2016}  Calculated ground-state minimum energy paths for the three pathways are shown in Fig. S1 of the supplemental material. An important finding of our calculations is that metal-surface adsorption increases the relative energy difference between E and Z isomers due to a preferential adsorption of the E-Ab state. This additional stabilisation arises from a strong contribution due to dispersion interactions between the phenyl groups of the flat-lying E-Ab molecule and the substrate. As shown in Fig.~\ref{fig-results-ab-rotation} and further summarised as $\Delta$E(E-Z) in Table 1, for the three coinage metals (Au, Ag, Cu), this effect follows the trend of substrate reactivity with Cu showing the strongest effect and Au the weakest. 

The most striking finding is that, along the  three reaction pathways, only the transition state along the dihedral rotation (TS) is significantly modified, which is why, in the following, we only focus on this pathway. As shown in Fig.~\ref{fig-results-ab-rotation}, the sizeable gas phase barrier of 1.09~eV when measured from the Z-Ab state, is strongly reduced upon adsorption. This effect is stronger for more reactive substrates and, in the case of Cu(111), reaches a point where the barrier vanishes completely and the Z-Ab state seizes to be a (meta)stable geometry. This finding is corroborated by the fact that no experimental evidence of a Z-Ab state on Cu(111) exists~\cite{alemani2008}. Even for a  Ag(111) substrate, the remaining barrier of 0.04~eV will likely provide sufficient stabilisation of the Z-Ab state for experiments at finite temperatures. Already from this data it becomes evident that Ab can not act as a functioning molecular switch on any metal surface that is more reactive than Au(111), with experiments failing to find any evidence of switching even on more reactive Au facets such as Au(100)~\cite{alemani2008}.

As pointed out in ref.~\cite{Maurer2012}, the mechanism behind this rotational barrier reduction lies in charge-transfer between the metal and the adsorbate in the rotational transition state geometry. Upon breaking the N-N double bond, the metal stabilises the transition state by directly donating electrons to it. The electronegativity of the metal or its ability to accept electrons is inversely proportional to its reactivity and to the position of the Fermi level with respect to the molecular states. From Au to Cu, the barrier consistently decreases with decreasing electronegativity or increasing reactivity of the underlying metal. Thus, despite the potential existence of an efficient electron- or light-induced E-to-Z isomerisation, a thermal Z-to-E back-reaction becomes more and more favourable.

The electronegativity of the substrate, therefore, represents a central design parameter for metal-supported molecular switches. A potential strategy to enable the light-induced isomerisation of Ab is consequently to increase the electronegativity of the underlying substrate or, equivalently to reduce the availability of surface electrons to stabilise the transition state. However, it should be stressed that the energy landscape is highly sensitive to substrate electronegativity as already the step from a Au to a Ag surface destroys the molecular switching ability and, for adsorption on Cu(111) at higher coverages, even dissociation of the N-N bond has been observed.~\cite{Willenbockel2015} A subtle increase in electronegativity can, for example, be achieved by depleting electrons from the surface with coadsorbates that are strong electron acceptors.  In the following, we will study a different avenue of tuning the switching ability, namely by molecular functionalisation with bulky spacer groups.

\begin{table*}
	\begin{center}\caption{\label{tab-results-Ab-TBA}  Central geometric parameters and relative energetics of Ab and TBA conformers in gas phase and adsorbed at coinage metal surfaces. In the case of TBA, $z$ refers to the vertical adsorption height of the lowest-lying N atom.  In the case of Ab, $z$ refers to an average of the two N atoms. $\Delta$E(E-Z) refers to the relative energy difference between E and Z conformer and $\Delta$E(TS-Z) between transition state and Z conformer.}
		\begin{tabular}{ccccccccc} \hline
			& \multicolumn{2}{c}{E} & \multicolumn{2}{c}{TS} &\multicolumn{2}{c}{Z} &  &   \\ \hline
			system & z & $\omega$ & z & $\omega$& z & $\omega$ & $\Delta$E(E-Z)  &  $\Delta$E(TS-Z) \\ 
			& \AA & deg & \AA  & deg & \AA & deg & \multicolumn{2}{c}{eV}  \\ \hline
			Ab  & - & 180 & - & 90 & - & 12 & 0.58 & 1.09  \\
			Ab@Cu(111)  & 2.04 & 169 & 1.85 & 90 & 1.78 & 20 & 1.22$^*$ & -  \\
			Ab@Ag(111)  &  2.52 & 173 & 2.08  & 90  &  2.10 & 50 & 0.77 & 0.06 \\
			Ab@Au(111)  &  3.03 & 179 &   & 90 &  2.48 & 19  & 0.73 & 0.37 \\
			TBA & - & 179 & -  & 90 & - & 8.3 & 0.33 & 1.27  \\
			TBA@Ag(111) & 2.36 & 155 & 2.06 & 89 & 2.07 & 29.7 & 0.48 & 0.04  \\
			TBA@Au(111) & 2.67 & 166 & 2.00 & 90 & 2.31 & 8.7 & 0.46 & 0.48 \\ \hline
		\end{tabular}\\
	$^*$ The Z-Ab state on Cu(111) is not stable. For the purpose of this comparison it has been defined as the Ab molecule with an $\omega$ angle constrained at $20^{\circ}$.
	\end{center}
\end{table*}

\subsection{The role of molecule functionalisation}
\label{results-tba}

A key strategy of molecular design is the derivatisation of molecules with purpose-designed functional groups to control the strength of molecular interactions. The four tetra-\emph{tert.}-butyl legs in TBA have been designed with the intent to decouple the molecule from the metal surface and to enable the photo-induced molecular switching. Contrary to Ab on Au(111), which can only be switched via electron-tunnelling through an STM tip, TBA on Au(111) can indeed be switched with light.~\cite{cho2010,comstock2008} However, the original idea that this is purely due to spatial decoupling has been debunked by quantitative x-ray standing wave measurements (XSW) and equilibrium structures of E-TBA on Au(111) calculated with DFT, which show that the vertical adsorption height of the central di-nitrogen bond is not significantly altered.~\cite{McNellis2010,mcnellis2010a}  In fact, our PBE+vdW$^{\mathrm{surf}}$ calculations (see Table 1) find vertical adsorption heights of the nitrogen atoms for E-TBA and Z-TBA that are even lower than for E-Ab and Z-Ab. The strongest reduction can be found for E-TBA on Au(111). In this case, this is mostly due to an asymmetric adsorption of the di-nitrogen bridge with a difference in height between the two nitrogen atoms of 0.3~\AA{}. 
\begin{figure}
	\centering\includegraphics{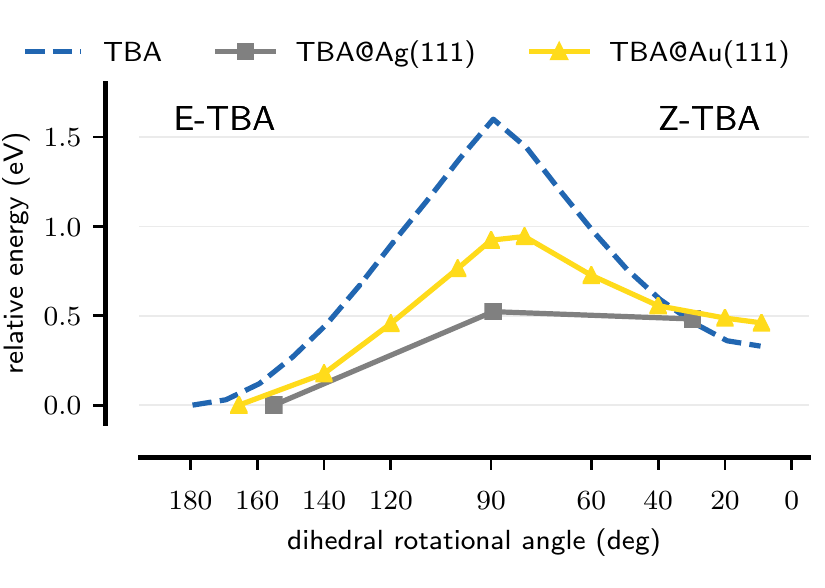}
	\caption{\label{fig-results-TBA-substrates} Minimum energy paths from E-TBA to Z-TBA along the dihedral rotation coordinate. The paths are shown for TBA in gas phase (blue line, dashed) and adsorbed to an Ag(111) (grey line, squares) and Au(111) (yellow line, triangles) surface.}
\end{figure}

As Fig.~\ref{fig-results-TBA-substrates} shows, the spacer groups in TBA also do not have any drastic effects on the ground state energy landscape. For TBA on Ag(111) and Au(111), the rotational barriers are strongly reduced compared to the free molecule and the remaining barriers compared to the Z-TBA state are comparable to those of Ab adsorbed on the respective surfaces with an almost vanishing thermal barrier for Z-TBA on Ag(111) and a barrier of 0.48~eV for Z-TBA on Au(111). This means that there are no thermodynamic arguments against the existence of a Z-TBA state on Au(111), whereas long-time thermal stability of a Z-TBA state on Ag(111) is highly improbable. Both of these assertions, as in the case of Ab, are in agreement with experimental reports on these systems.~\cite{comstock2007,alemani2008}

An interesting structural difference that can be observed is the change in $\omega$ angles for the equilibrium E and Z-TBA structures compared to adsorbed Ab structures. Whereas E-Ab adsorbs in an almost flat-lying geometry on the Ag(111) and Au(111) surfaces, E-TBA shows deviations of 25$^{\circ}$ and 14$^{\circ}$, respectively, from a perfectly flat arrangement with the phenyl rings and side groups distorted away from the surface. On the other hand, the Z-TBA structures show much more acute angles than their Ab counterparts when adsorbed on Ag(111) and Au(111) (see also Fig. S2 and Fig. S3 for molecular structures of the metal-surface adsorbed Ab and TBA equilibrium geometries). Despite these differences, the difference in angle $\Delta\omega$ that defines the reaction path from E to Z is almost the same for Ab and TBA on Ag(111) and Au(111), namely 123/125$^\circ$ and 160/157$^\circ$, respectively. However, the effective angle distance from the E-TBA state to the rotational ground-state barrier is reduced for E-TBA compared to E-Ab. On Au(111), this difference is 89$^{\circ}$ for E-Ab and 76$^{\circ}$ for E-TBA. In addition, the energy difference between the E and Z  states is significantly smaller when compared to Ab. 

This means that, in the context of the proposed light- or electron-driven mechanism that involves short-lived transient excited states, which provide a funnel for energy from hot electrons into vibrational degrees of freedom, less total energy has to be provided to make it from the E-TBA to the Z-TBA state compared to Ab and an effectively shorter path has to be passed to cross the barrier. The first is accomplished by a relative destabilisation of the E-TBA state compared to the rotational transition state and the Z-TBA state due to the bulky spacer groups. The latter arises from the distorted di-nitrogen center in the ground state structure of E-TBA. We  therefore note that, while bulky spacer groups do not geometrically decouple the di-nitrogen bridge from the surface, their destabilising effect on the E-TBA geometry will facilitate switching.

Whereas a clear picture arises for the switching ability of Ab and TBA on Ag(111) and Ab on Cu(111) and, by deduction, any other more reactive metal surface, the subtle differences between the ground-state potential energy landscapes and equilibrium structures of Ab and TBA on Au(111) do not provide an explanation for the difference in switching ability between the two. In order to understand the thermodynamic and kinetic differences that give rise to the photoswitching ability of TBA on Au(111), we therefore proceed to study the energetic landscapes of molecule-surface charge-transfer excitations proposed in the context of the on-surface isomerisation reaction.

\subsection{Excited state energy landscapes of transient molecular ions}
\label{results-excited-state}

The requirements of potential energy landscapes to be able to support a successful transient ion-based photoswitching mechanism, as the one described by Wolf and Tegeder,~\cite{Wolf2009} are somewhat different from the ones that apply to gas phase and solvent photochemistry. First and foremost, the transient ion formation has to be efficient, \emph{ie.} the transient ionic state has to have high overlap with high-density areas of the metallic density-of-state, such as the d-bands, so that electrons or holes can be efficiently transferred. The probability of switching due to this transient ion will be high if, within its short lifetime of few tens of femtoseconds,~\cite{campillo2000} it can efficiently transfer electronic excess energy from the vertical excitation that led to the ion formation to vibrational energy within the reactive degrees of freedom of the molecule that will drive the isomerisation. This will be the case if the equilbrium geometry exhibits strong forces upon ion formation. Whereas, in the context of an intramolecular excitation, changes to the excited state landscape upon surface-adsorption are to be avoided to retain the switching function, in the context of a transient ion formation, changes to the energy landscape of the cationic state induced by substrate variation or molecule functionalisation could be advantageous. To enable an efficient transient-ion-driven isomerisation, the energy landscapes of transient ion excitations need to show the following features: 
\begin{enumerate}[1.]
	\item a significant gradient towards the transition state  within the Franck-Condon (FC) region of the excited state PES,
	\item a barrierless path towards the ground-state transition state,
	\item and a kinetic excess energy (\emph{vide infra}) that exceeds the ground state barrier, but is not too large to increase the probability of immediate thermal Z-to-E backreaction.
\end{enumerate}

The natural starting point for an investigation of the adsorbed PES topology of ionic resonances are the cation and anion PESs of isolated Ab and TBA. Corresponding gas phase calculations have been reported by F\"uchsel \emph{et al.}~\cite{Fuchsel2006} and Leyssner \emph{et al.}~\cite{Leyssner2010} in the context of electron-tunneling- and light-induced Ab switching on metal surfaces. STM-induced switching can be triggered with negative and positive bias voltage. Therefore, electron removal ('cationic resonance') and electron attachment ('anionic resonance') to the molecule are both viable mechanisms, whereas in the case of light-induced switching of Ab and TBA the possibility of an anionic resonance as dominating excitation mechanism has been excluded by experimental considerations.~\cite{Hagen2007, Wolf2009} 

The above mentioned studies report DFT-based anionic and cationic potential energy surfaces of Ab and TBA along rotation and inversion. Just as in the ground-state, these PESs show two meta-stable states at very similar positions, although slightly shifted towards smaller $\alpha$ bending angles and $\omega$ dihedral angles, closer to the mid-rotation point. Correspondingly, both ionic states show significant gradients at the FC structures of E and Z-Ab and TBA. Additionally upon electron removal or attachment, the rotation and inversion barriers are significantly reduced, although still present. In general, the ionic surfaces are less corrugated and specifically the cationic PES is much more shallow. It can therefore be argued that in both ionic molecular states, the prerequisites to a substrate-mediated excitation mechanism are in principle given and if only minimal changes to these PESs occur upon metal-surface adsorption, light-induced switching along the Wolf-Tegeder mechanism is feasible.

\begin{figure*}
	\centering\includegraphics[width=\textwidth]{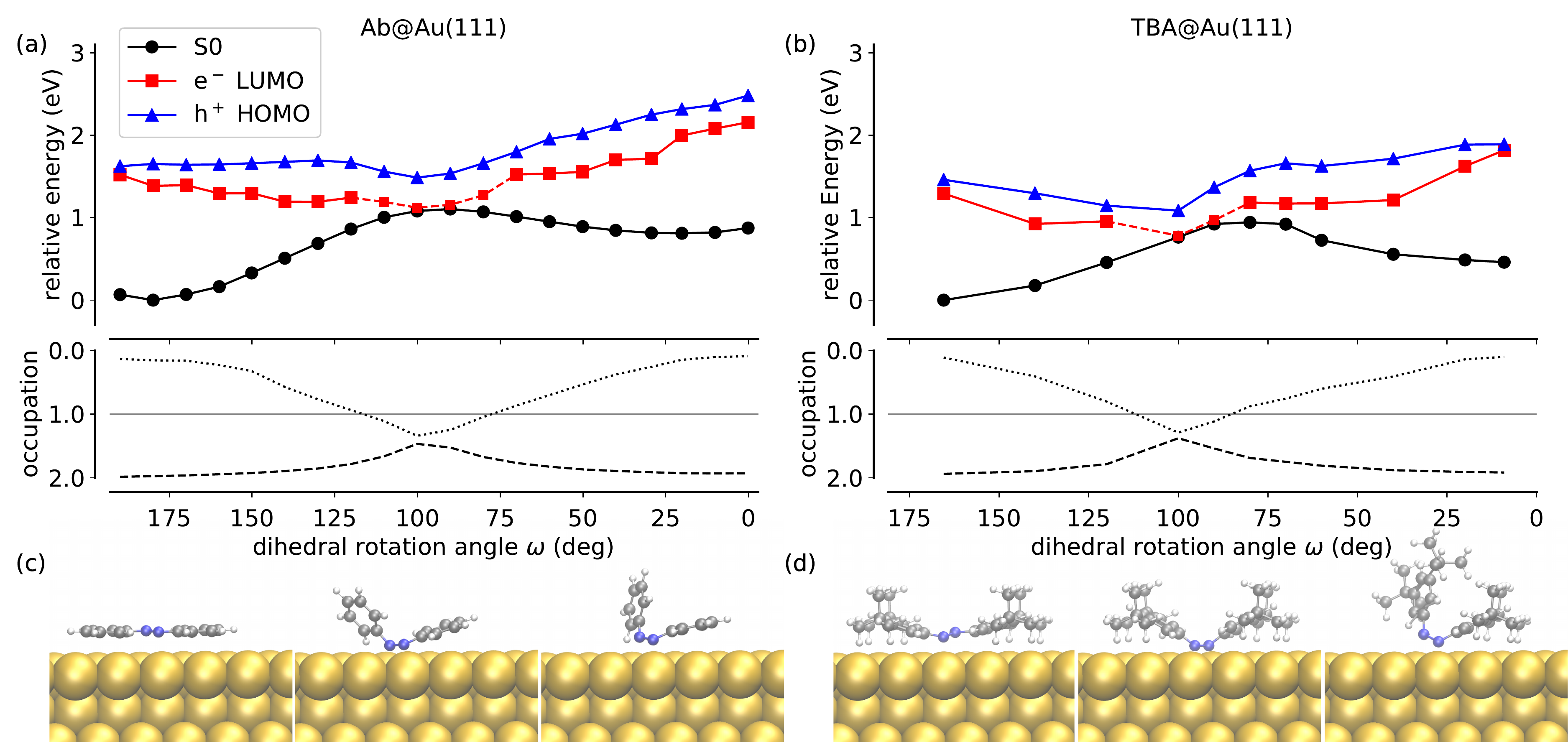}
	\caption{\label{fig-Ab-TBA-CT} (a) and (b) Upper panel: Relaxed minimum-energy paths of Ab (a) and TBA (b) adsorbed on Au(111) along the dihedral rotation coordinate. Shown are the ground-state PES (black), an anionic resonance where an electron has been added to the LUMO of the molecule ($e^{-}$, red), and a cationic resonance where an electron has been withdrawn from the HOMO of the molecule ($h^{+}$, blue). Points marked with dashes for the anionic resonances are calculations where the excitation constraint in le$\Delta$SCF-DFT could not be fully satisfied due to the already existing occupation of the LUMO in the ground-state (see the main text for more details).  Lower panel: Ground-state occupations of the molecular HOMO (dashed line) and the molecular LUMO (dotted line) along the dihedral rotation pathway. These occupations have been calculated using the orbital projections that underlie the le$\Delta$SCF-DFT method.~\cite{Maurer2013}. (c) Side view of Ab equilibrium and transition state structures along the isomerisation pathway. The order follows: E-Ab, TS-Ab, Z-Ab (d) Side view of TBA equilibrium and transition state structures along the isomerisation pathway. The order follows: E-TBA, TS-TBA, Z-TBA}
\end{figure*}

Figure \ref{fig-Ab-TBA-CT} presents the calculated cationic (in blue) and anionic (in red) resonance states for Ab and TBA adsorbed on Au(111) along the minimum energy path of dihedral rotation. As described in the previous sections, the ground state PES along this pathway (shown in black) exhibits a very similar topology for both molecules, which derives from the similar adsorption-induced changes that occur along this pathway. Both E and Z states of Ab and TBA are dominantly physisorbed, which is reflected in a net occupation of the molecular LUMO that is close to zero and a net population of the molecular HOMO that is close to full occupation (see occupation panels in Fig.~\ref{fig-Ab-TBA-CT}). These occupations are calculated by projecting the  molecular orbitals of the isolated molecule from the band structure of the slab with the adsorbed molecule and integrating the corresponding occupations up to the Fermi level.~\cite{Maurer2012,Maurer2013} In a previous study, we have compared this charge analysis measure with other more established methods and have found it to provide a description of net charge flow between that of the Hirshfeld and the Bader methods.~\cite{Mueller2016}  The loss of occupation in the HOMO and the gain of occupation in the LUMO around the rotational transition state is a clear sign for a covalent bond formation via a surface-to-molecule donation into the LUMO and back-donation of charge from the HOMO into the surface. 

The here presented excitation energies for the cationic and anionic resonance represent the lowest possible vertical transitions of electrons between molecular states and metal states at the Fermi level based on the equilibrated ground-state geometries. Corresponding excitations can also happen at higher photon energies albeit with more excess energy. For both molecules, we find the cationic resonance at a higher excitation energy than the anionic resonance, as expected considering that the IP of the isolated molecule is much larger than the EA. At the E-Ab geometry, the corresponding energies (h$^+$: 1.65~eV, e$^-$: 1.39~eV) are close to the molecular resonance energies measured with 2PPE (HOMO: 1.77~eV, LUMO: 1.68~eV, absolute energy difference from the Fermi level~\cite{Bronner2012}) with the same qualitative trend of HOMO$>$LUMO. The anionic resonance shows a larger discrepancy, which could be a reflection of the poor description of unoccupied states with semi-local DFT and possibly also due to the difference between the high-coverage situation in experiment and the intermediate-coverage as described by our calculations. 

For both molecules, the anionic resonance (shown in red in Fig.~\ref{fig-Ab-TBA-CT}) clearly satisfies the above mentioned criteria. The path from E-Ab and E-TBA towards the transition state is barrierless and shows a moderate gradient. The nominal excitation energy, not assuming any dissipation during the transition, is sufficient to overcome the barrier. We should note that the description of this excitation around the rotational transition state is limited in the le$\Delta$SCF-DFT method as the LUMO of the molecule is already more than half filled between 125 and 75$^{\circ}$ (dashed line). This means that we cannot excite a full electron into the LUMO and therefore cannot fully satisfy the occupation constraint. This is, however, also an indication that the lifetime of such a state will be extremely short. Considering also the experimental assessment that suggests that the isomerisation is not triggered by an anionic resonance, we will therefore focus on the cationic resonance state in the following.


For the cationic resonance states of TBA and Ab on Au(111), more differences are visible. The $h^{+}$ surface for both molecules shows a minimum close to the rotational transition state. However, the E-TBA FC region exhibits a significant gradient towards the transition state, whereas the E-Ab FC region shows an extended plateau across a wide range of $\omega$ values. The electronic excess energy, \emph{ie.} the energy difference between the vertical excitation energy and the ground state barrier, is very similar for both systems (0.55 and 0.52~eV for Ab and TBA on Au(111)). Therefore the cationic resonance along rotation satisfies all above defined criteria for transient-ion-mediated switching for TBA on Au(111). The cationic resonance for Ab on Au(111) instead satisfies criteria 2 and 3, but violates criterion 1. This means that, while in principle possible, the insufficient coupling of molecular vibrations with the transient cation state, could prevent sufficient energy to be transferred to the molecular vibrations within the short lifetime of the excitation. This then would rationalize why, contrary to TBA on Au(111), no light-induced switching has been observed for Ab on Au(111), whereas E-to-Z switching can be induced with sufficiently high STM electron tunnelling currents. 

The excited-state topologies calculated for TBA do thus support the Wolf-Tegeder mechanism in several aspects. The transition from E to Z-TBA involves almost exclusively rotational motion around the central dihedral angle $\omega$, with the exception of a small, almost isoenergetic twisting of the phenyl rings when lifting off the phenyl ring at dihedral angles between the transition state and Z-TBA. The PES topology of a cationic resonance in the HOMO that is connected with this motion represents a fairly steep, barrierless descent towards the mid-rotation point. At this point, if not earlier, increased coupling with the substrate is likely to quench the already short-lived excitation and a corresponding motion will have a higher probability for crossing the ground-state barrier than for a momentum inversion to occur. The smaller gradient around the Z-TBA FC region and the correspondingly reduced vibrational activation of the rotational motion rationalise the lack of experimentally found Z-to-E TBA photoswitching. Furthermore, the strong vibrational dependence of the switching rate and the termal Z-to-E isomerisation that have been observed experimentally,~\cite{hagen-phd} are also consistent with the potential energy path that has been found here. A thermal activation of the dihedral angle, corresponding to a wagging motion of the molecule perpendicular to the surface, will strongly facilitate a corresponding switching upon photo-excitation.

Up to this point, we have only considered isomerisation along the dihedral rotation around the central di-nitrogen bridge. The literature around light-induced isomerisation in the gas phase discusses in detail the (co)existence of different isomerisation pathways that also include an inversion around one of the two central bending angles. For the sake of completeness, we note that cationic and anionic resonance excitations along such an inversion do not provide energy landscapes that satisfy the above criteria (see Fig. S4 in the supplemental material). Both, cationic and anionic resonance, show a topology similar to the free molecular cation and anion states with a barrier between the E-Ab and Z-Ab states.  It should be noted that the  steric hindrance due to bulky spacer groups of TBA will significantly impede  the isomerisation along the bending angle inversion, leaving dihedral rotation as the only viable pathway for E-to-Z isomerisation. Counterintuitively, the introduction of large functional side groups, as in TBA,  strongly decreases the number of degrees of freedom that can participate in isomerisation. As such, the here presented findings would suggest that, at least for the light-induced on-surface isomerisation, dihedral rotation is the dominant isomerisation pathway. This is contrary to a suggestion by Comstock \emph{et al.} that inversion must be the dominant isomerisation route.~\cite{Comstock2010}

Despite the apparent failure to stabilise a Z-Ab or Z-TBA state on Ag(111), it can be insightful to compare the effect of different substrates on the ionic PESs. Comparing the excited states along the rotation for Ab on Au(111) and Ag(111) (see supplemental material, Fig. S5), strong changes to the cationic and anionic resonance PESs are apparent. The corresponding paths of cationic and anionic resonance states along rotation are strongly modified, as compared to the gas phase or  to the molecule adsorbed on a Au(111) surface. Owing to the strong adsorption, almost everywhere along the path the LUMO is more than half-filled in the ground state and therefore the anionic resonance excitation constraint in le$\Delta$SCF can not be fully satisfied. Correspondingly, the anionic resonance state is systematically downshifted and coincides with the ground state over large portions of the rotational pathway. More importantly, the cationic resonance state is systematically shifted upwards on Ag(111) as compared to Au(111) and appears more corrugated with a pronounced minimum at the transition state. At both FC regions of E and Z-Ab on the cationic resonance state, there is a significant gradient towards the mid-rotation point. Quite ironically, whereas Ab adsorbed on Ag(111) nominally fulfils the necessary criteria to efficiently activate nuclear motion upon excitation, it fails to sufficiently stabilise a Z-Ab conformer. In contrast, in the case of Ab on Au(111) sufficient stability is accompanied by highly inefficient vibronic energy transfer. The inspection of the static PESs therefore suggests that both isomerisation processes can, at best, be highly inefficient.

The excited state PES differences between Ag and Au can be understood in terms of the differences in electronegativity of the underlying surfaces. The electronegativity of the Au surface is higher than that of the Ag surface.  Therefore, electrons are harder to detach from the Au surface than from the Ag surface, or in other words, adding electrons to a gold surface is connected to a smaller energetic penalty than for the case of silver. When inducing a cationic resonance state, an electron from the molecule is transferred to the substrate. For the above reasons, the energy that is necessary to do this is higher for Ag than for Au. In the case of an anionic resonance this situation is reversed. The more electronegative a surface, the higher the energetic penalty to withdraw electrons from it. Correspondingly, the energy associated with the anionic resonance on the molecule is higher for E-Ab on Au(111) than on Ag(111). This, however, does not hold for the resonances at the transition state. In both cases, the chemical bond to the surface efficiently transfers electrons from the substrate to the LUMO of the molecule. Correspondingly, the ground state is already, to some extent, an 'anionic resonance' state compared to the isolated molecule; the ground state and the $e^{-}$ state are energetically equivalent. Detaching an electron at the mid-rotation point from the HOMO, effectively yields an $n\rightarrow\pi^*$ excited Ab molecule. The corresponding excitation energy at the transition state, due to the orbital degeneracy and the conical intersection with the ground state, is zero in the gas phase. The minimum observed in the cationic state at the transition state for the adsorbed molecules stands in contrast to the barrier exhibited at this point for the gas phase molecules.~\cite{Fuchsel2006} This minimum appears because of the significant mixture of the cationic state with the n$\rightarrow\pi^*$ (S1) intramolecular excitation that stems from the molecule-surface charge-transfer in the ground-state. Correspondingly, at this point, the hybridization with the surface completely inverts the topology of the cationic resonance state and enables an otherwise infeasible substrate-mediated light-induced isomerisation process.

\section{Discussion and Outlook}
\label{conclusions}

The combined picture of ground- and excited state energy landscapes for Ab and TBA on Ag(111) and Au(111), in combination with previously presented intramolecular excited state landscapes,~\cite{Maurer2013} provides a clear rationalisation of the proposed switching mechanism in terms of the choice of substrate and the molecule functionalisation, as well as their effect on the underlying molecule-surface chemistry.

Reactive metal surfaces will not be able to sustain light- or electron-induced E-Z switching around covalent double bonds. This apparently includes noble surfaces such as silver and anything more reactive. The reason for this is an unbalanced stabilisation of different geometries along the reaction pathway, where the transition state and the E-Ab state are strongly stabilised with respect to the Z-Ab state by surface-molecule charge transfer and dispersion interactions, respectively. The effect is a loss of stability of the Z-Ab state in the ground state. 

In addition to a lack of ground state stability, such reactive substrates that easily donate electrons to adsorbates will also suppress anionic resonances to the LUMO and shift cationic resonances to high energies, which equivalently means reduced overlap with the d-bands. For Ab on a more inert substrate such as Au(111), due to the high substrate electronegativity, the ionic excitations are sufficiently separated from the ground state and the ground-state stability is such that, in principle, a bistable molecule with an E and a Z state can be supported. Nevertheless, surface adsorption (hybridisation and image charge interactions) strongly modifies the topology of the ionic resonance states compared to the isolated molecule with the effect that the coupling to molecular vibrations for the E-Ab state is minimal. Whereas simple UV light exposure will not be able to transfer sufficient momentum to molecular motion, inelastic electron tunnelling can still transfer sufficient kinetic energy to achieve an isomerisation to the Z-Ab state.

The bulky spacer groups in TBA have a much more faceted effect on the potential energy landscape and switching ability than previously considered. The additional side groups do not achieve a chemical decoupling of the central di-nitrogen bridge from the surface, however they distort the E-TBA conformation and effectively destabilise it compared to all other structures. As such they contribute to a more balanced adsorption along the reaction pathway. Yet, the charge-transfer at the rotational transition state and the dipole-image charge interaction in the Z-TBA state are still sufficiently large to modify the cationic resonance landscape compared to adsorbed Ab. As a result, the excited-state PESs are shifted downwards for the rotational transition state and the Z-TBA isomer: what was a plateau for the cationic resonance state in the case of E-Ab, for E-TBA, shows a gradient that evidences the ability to transfer electronic excess energy into vibrational motion towards the barrier along a path that is significantly shorter than it is for E-Ab. These effects compound to a much higher reaction probability, while not significantly affecting the level alignment and the rate of ion formation.

The above presented results may also help to settle the controversy over the prevalence of rotation or inversion motion during photo-isomerisation. Comstock {\em et al.}, based on their STM experiments, have proposed that a dominance of an inversion-based mechanism is likely.  The authors base this interpretation on the observation of a selection rule to photo-switching that applies differently to different chiral island domains of TBA on Au(111). They rationalize this effect by an initially planar linearisation of one CNN bond angle $\alpha$ and two subsequent channels towards the Z-TBA structure. In that way, E-TBA of one racemic type can convert into a left-handed and a right-handed Z-TBA molecule by changing the handedness of the di-nitrogen bridge. As a second argument, such a pathway would additionally maximise the dispersion interaction of both phenyl rings with the surface as much as possible during the isomerisation process.

The here presented  ground- and excited-state potential energy landscapes paint a different picture. Whereas surface adsorption strongly affects the rotational pathway, minimal effects are apparent for the isomerisation along bending angle inversion. Both intramolecular excitation or transient ion formation would lead to barriers in the excited state and an insufficient driving force towards isomerisation. The corresponding ground state barrier that has to be overcome is considerably higher for inversion than in the case of rotation. Previously found low-lying vibrational modes contribute to vertical wagging modes that can initiate a rotational motion, whereas no such low-lying modes where found that contribute to asymmetric bending angle inversion.~ \cite{McNellis2010} On the basis of our results, dihedral rotation provides the most likely pathway to isomerisation. This does, however, not mean that some bending angle motion might not contribute to the dynamical process and the intramolecular vibrational energy distribution. The chiral selectivity that forms the argument for an inversion-based mechanism can be understood by taking a look at the dynamics of the mechanism for the molecules in solvent. Weingart \emph{et al.}~\cite{weingart2011} have found that also dihedral rotation can generate two different pro-chiral Z-Ab variants by either clockwise or anti-clockwise rotation. This finding is based on the fact that the dominant motion during isomerisation is carried out by the light-weight di-nitrogen bridge rather than the side groups. In the case of surface-adsorbed molecules, that could mean that during a wagging motion of the molecule, an asymmetric vertical rotation of the two nitrogen atoms (much like the motion of pedals on a bicycle when viewed from ahead) may change the handedness of the molecule. This assertion is supported by the recent finding of such motion for an Ab tripod structure on Ag(111).~\cite{Scheil2016}


Our findings show that the le$\Delta$SCF-DFT approach, at least qualitatively, accounts for a range of important effects that shape on-surface excited states, including electrostatic stabilisation via interaction between excited-state dipoles and the substrate image charge, as well as hybridisation and charge-transfer in chemisorbed geometries. Although no fully  quantitative description can be expected from an approximate scheme like le$\Delta$SCF-DFT, the presented excited state topologies offer a convincing rationalisation of the experimental findings for the molecular switching of Ab and its derivatives on the close-packed coinage metal surfaces. One of the current limitations of our implementation of le$\Delta$SCF-DFT is the inability to calculate analytical forces in the excited state. This will be key to ensure its widespread usage and its successful application for complex and high-dimensional light-driven dynamics at surfaces. With the advent of the machine-learning-based construction of high-dimensional energy landscapes and more from {\em ab initio} data,~\cite{Behler2007b,Kolb2017,Rupp2018}  this limitation could be alleviated. The authors thus see a clear remit for this efficient and approximate excited-state methodology to tackle challenging problems in photocatalysis and nonadiabatic dynamics at surfaces and nanoparticles. 

\section{Conclusion}
\label{conclusion}

We have performed state-of-the-art dispersion corrected DFT and linear expansion $\Delta$SCF-DFT calculations to study the potential energy landscapes of the ground state and molecule-surface charge-transfer excited states of Azobenzene and its derivative TBA adsorbed at coinage metal surfaces. In the context of light- and electron-tunnelling-induced molecular switching of Ab, we investigated the traditional experimental parameters of functional interface design, namely the choice of substrate and its reactivity, and the molecular functionalisation. Our findings provide evidence of the drastic and somewhat unexpected effects that these parameters have on the energy landscape and the expected reactivity of the two molecules.

Our findings fully support a previously proposed mechanism based on the transient formation of a cationic molecular resonance due to light exposure or electron-tunnelling. We can clearly discern the effects that the change of substrate and the introduction of bulky spacer groups have on the energy landscapes, the molecule-surface interaction and the reactivity. The differences in the ground and excited-state potential energy landscapes that our calculations expose, clearly rationalise why TBA on Au(111) can be switched via light, whereas Ab on Au(111) can only be switched via inelastic electron tunnelling through an STM tip. 

In addition to providing mechanistic insight into the photochemical reactivity of a prototypical functional interface and a differentiated picture of the effect of traditional experimental design strategies, our calculations provide evidence of the potential utility of the le$\Delta$SCF-DFT method for a diverse range of chemical challenges associated with charge-transfer in condensed matter.

\ack
Funding through the Deutsche Forschungsgemeinschaft (DFG) is gratefully acknowledged. Computer resources for this project have been provided by the Gauss Centre for Supercomputing/Leibniz Supercomputing Centre under grant id: pr63ya and the Garching Supercomputing Center of the Max-Planck Society.



\section*{References}


\begin{thebibliography}{10}
\bibitem{comment1}
	The employed vacuum slab, as dictated by computational resources, was 22~\AA{}
	as measured from the molecule to the image of the lowest metal layer,
	therefore a higher vertical distance would lower the excitation energies
	again.
\end{thebibliography}

\providecommand{\newblock}{}

\end{document}